\documentclass[12pt]{iopart}

\usepackage{color}
\usepackage{amssymb}
\usepackage{graphicx}   
\usepackage{subfigure}
\usepackage{bm}

\begin{document}     
	\title{Nickel coated carbon nanotubes in aluminum matrix composites: A multiscale simulation study} 
	\author{Samaneh Nasiri$^1$, Kai Wang$^2$, Mingjun Yang$^2$, Qianqian Li $^3$, Michael Zasier $^1$$^,$$^2$$^,$$^4$ }
	
	\address{$^1$WW8-Materials Simulation, Department of Materials Science, Friedrich-Alexander Universit\"at Erlangen-N\"urnberg, 90762 F\"urth, Germany} 
	\address{$^2$School of Materials Science and Engineering, Southwest Petroleum University, Chengdu, Sichuan, PR China}
	\address{$^3$Department of Aerospace Engineering, Imperial College, Prince Consort Road, London SW7 2BZ, UK}
	\address{$^4$School of Mechanics and Engineering, Southwest Jiaotong University, Chengdu, Sichuan, PR China}
	\ead{samaneh.nasiri@fau.de}
	\begin{abstract} In this work we use density functional theory (DFT) calculations to benchmark empirical potentials for the interaction between nickel and sp$^2$ bonded carbon nanoparticles. These potentials are then used in order to investigate how Ni decorated or coated carbon nanotubes (CNT) affect the mechanical properties of Al/CNT composites. In particular we look at the pull-out behaviour of pristine as well as Ni-decorated and Ni-coated CNT from an Al matrix. Our result shows that Ni coating may produce an extended interface (“interphase”) where a significant amount of energy is dissipated during CNT pull-out, leading to a high pull-out force. We also demonstrate that surface decorated CNT may act as efficient nano-crystallization agents and thus provide a novel strengthening mechanism not previously discussed in the literature. We discuss our results in view of  promising approaches for engineering CNT-metal interfaces such as to achieve high strength metal-CNT composites. 
	\end{abstract}
	
	\maketitle
\section{Introduction}
\label{intro}

Carbon nanotubes (CNT) possess excellent mechanical properties in terms of axial elastic stiffness and rupture strength \cite{demczyk2002-J.Mater.Sci.Eng.A,yu2000-Science,Nasiri2016}. It is therefore an obvious question whether these properties can be harnessed for structural applications, for instance, by using CNT as fillers in nanocomposites with low-melting lightweight metals such as Al or Mg as matrix. The benefits, in view of mechanical properties, of embedding carbon nanoparticles into metals are  potentially huge \cite{li2009-compos.sci.technol,george2005-Scripta.Materialia,bakshi2010-Int.mater.rev}. To achieve these benefits, homogeneous dispersion of CNT in the metal matrix and strong interfacial bonding are essential factors: nanoparticle agglomerates with weak bonding into the surrounding metal matrix might act as flaws and thus deteriorate, rather than improve, the mechanical properties of composites \cite{li2010-compos.sci.technol}. However, good dispersion and strong interfacial bonding are hindered by the low affinity of CNT to Al or Mg. Interface engineering approaches, such as coating or decorating the CNT with metal nanoparticles, may help to overcome this problem and to allow efficient fabrication of lightweight metal-CNT nanocomposites \cite{kim2009-Synthetic.Metals,guo2002-J.Mater.Sci.Eng.A}. In addition to improved wetting by lightweight metals, coating or surface decoration of CNT with metals may provide further benefits. If instead of a continuous coating layer, isolated metal nanoparticles are deposited on the CNT surface, then these nanoparticles may serve a dual purpose: Discrete metal nanoparticles on the CNT surface can help to prevent CNT agglomeration by the simple means of acting as geometrical ’spacers’. This effect is well known in the context of graphene where decoration of exfoliated graphene sheets with Pt nanoparticles was shown to prevent face-to-face aggregation of the sheets \cite{si2008-chem.of.Mat}. Attaching metal nanoparticles rather than continuous metal coatings to CNT may have further benefits. If a metal such as Ni is used that bonds well to sp$^2$ carbon, the interfaces between the nanoparticles and the CNT may be comparatively strong. The same is true for the nanoparticles themselves which have high  mechanical strength because they are too small to contain dislocations.  Such nanoparticles might therefore act as a ’nano-rivets’ which enhance interfacial shear stress transfer once the decorated CNT is embedded in a metal matrix. 

\begin{figure*}
\begin{center}
	\resizebox{0.99\textwidth}{!}{%
		\includegraphics{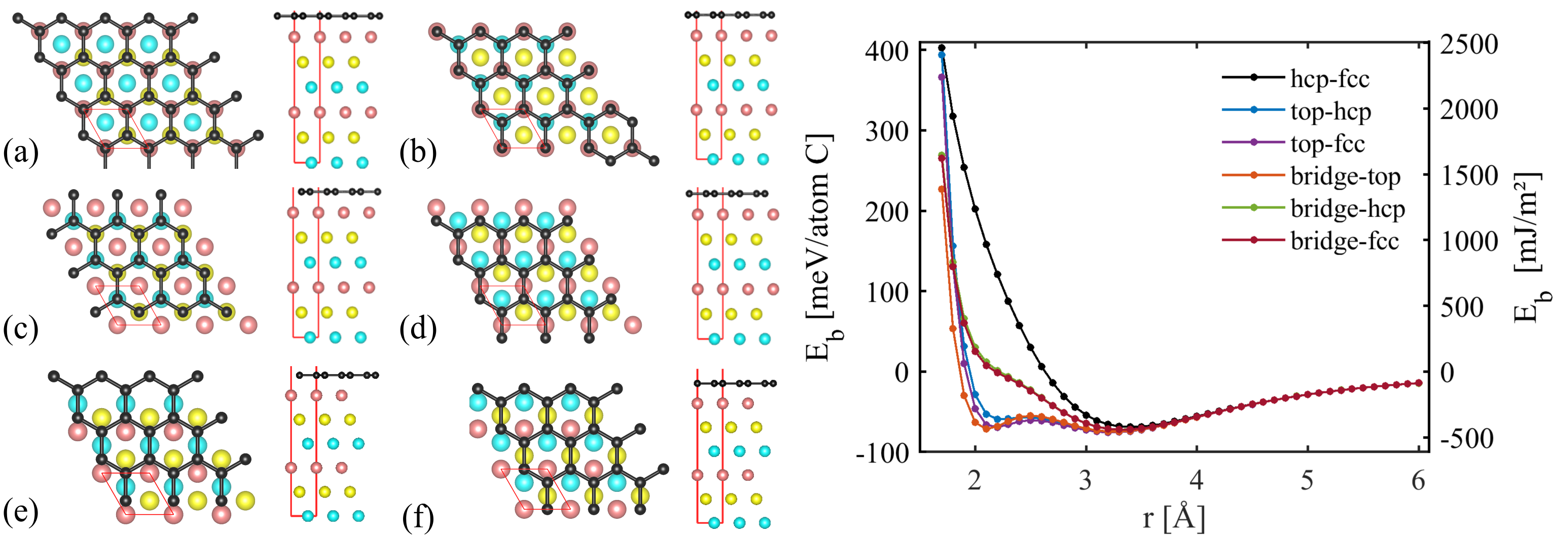}}
	\centering
	\caption{Ni(111)-graphene binding energy (per carbon atom) as a function of distance for different interface configurations; (a): top-hcp, (b): top-fcc, (c): hcp-fcc, (d): bridge-top, (e): bridge-hcp, (f): bridge-fcc .}
	\label{fig:DFTconfig}  
\end{center}    
\end{figure*}

There exist several MD simulation studies that investigate the mechanical behavior of metals with embedded CNTs. Silvestre et. al. \cite{Silvestre2014_CST} use MD to simulate the compressive buckling of CNT embedded into an Al matrix, and Choi et. al. \cite{Choi2016_CB} investigate the behavior of a CNT-Al system under tensile loading. The possibility of improving interfacial properties in CNT-metal systems by Ni coating of CNTs has also been investigated. Song et. al. \cite{Song2010_CMS} use MD to compare the simulated pull-out behavior of Ni coated and uncoated CNTs from an Al  matrix, and Duan et. al. \cite{Duan2017_PhysicaE} conduct a similar MD investigation with Cu as matrix material. These investigations show that Ni coating may significantly enhance the CNT pull-out force. In these works, the bond-order potential of \cite{Shibuta2007_CMS} is used for describing Ni-Carbon interactions. This potential is combined with EAM potentials for the Ni-Al or Ni-Cu alloy systems, and a Brenner \cite{Song2010_CMS} or AIREBO \cite{Duan2017_PhysicaE} potential for the C-C interactions. In physical terms such a combination appears problematic: One assumes that Ni atoms located at the Ni-CNT interface and chemically bonded to Carbon atoms interact with other metal atoms (Ni and Al or Cu) in exactly the same manner as if they were located at a free surface. We cannot easily see the physical justification for this assumption.

In the present paper we therefore address the problem in a multiscale framework. We first perform density functional theory (DFT) calculations to establish the energetics of interfaces between Ni and sp$^2$ bonded carbon. To this end we consider planar Ni-graphene interfaces which allow for the use of periodic supercells. The results of these calculations are reported in Section 2. They provide us with a clearer understanding of the bonding situation at Ni-graphene interfaces, which turns out to be unusual as a physisorbed state (bonded by Van der Waals interactions) co-exists with a chemically bonded state of approximately equal energy. In addition, we obtain reference data for assessing the performance of different interaction potentials in describing not only the interface energy but also the interface energy 'landscape', i.e. the changes in interface energy that occur if the graphene is displaced with respect to the Ni. We then perform, in Section 3, MD simulations to investigate the behavior of Ni coated CNT embedded in Al. In these simulations we consider two alternative descriptions of the Ni-CNT interface: (i) we use the DFT data of Section 2 to parametrize a Morse potential for describing Van der Waals like interactions between Ni and CNT in the physisorbed interface state, and (ii) we consider an alternative description of the ternary Ni-C-Al system in terms of reactive force fields (ReaxFF) that better account for the bonding situation at a chemisorbed Ni-C interface. The results obtained by these two approaches are compared. A general discussion of our results is given in Section 4.

\section{Density-functional calculations}
\label{sec:2}

To understand the interaction between Ni coating and CNT we use, as a reference system, a planar graphene sheet interacting with a Ni slab. This system -- which can be considered the limiting case of a fully coated single-walled CNT (SWCNT) of infinite diameter -- allows for periodic boundary conditions to be used in DFT calculations. Specifically, we consider monolayer graphene adhering to a six-layer Ni(111) slab. We studied different configurations of the graphene relative to the Ni slab as shown in Figure \ref{fig:DFTconfig}. A vacuum gap of at least 10 $\mathrm{\AA}$ was imposed in the slab-perpendicular $z$ direction when applying periodic boundary conditions. First principles calculations were performed using the Quantum Espresso package version 6.1 \cite{giannozzi2009-J-Phy-Con-Mat}, and PBE-based projector augmented wave \cite{blochl1994-PRB} (PAW) potentials were adopted. Previous studies \cite{graziano2012-J-Phy-Con-Mat,silvestrelli2015-PRB,munoz2016-Theory-Chem-Accounts} indicated that the optB88-vdW \cite{becke1988-PRA} functional produced reasonable results which were very close to experimental values, so we used this  functional in the DFT calculations. The energy cutoff was set to 240 Ry, a Methfessel-Paxton smearing of 0.01 Ry was used for the electronic convergence, and the convergence for self-consistency calculations was less than 10$^{-9}$ Ry. The Monkhorst-Pack $k$-points grid was \((8 \times 8 \times 1)\) for all configurations during calculations. Since our main focus is to obtain a general understanding of the interface energetics and binding situation, and to obtain data for parameterizing a phenomenological interaction potential, we did not carry out structural relaxation. Instead we simply displaced the graphene sheet rigidly with respect to the Ni(111) slab without geometry relaxation of either the Ni slab or the graphene. We evaluated the binding energy per carbon atom as

\begin{eqnarray}
E_{\rm b} = [E_{\rm tot} - (E_{\rm Ni} +E_{\rm G})]/n 
\label{eq1} 
\end{eqnarray}          

where $n$ is the number of carbon atoms in the graphene sheet, \(E_{\rm tot}\) is the total energy of the Ni-graphene system, \(E_{\rm Ni}\) represents the energy of the isolated six-layer Ni(111) slab and \(E_{\rm G}\) the energy of the free-standing graphene sheet.

As demonstrated in Figure \ref{fig:DFTconfig} we find two types of energy minima. A physisorption minimum is found at a distance of about 3.3 $\mathrm{\AA}$ between graphene and Ni slab. This energy minimum is separated by an energy barrier from a chemisorption minimum at a distance of about 2.2  $\mathrm{\AA}$. Both types of energy minima are of approximately equal depth. This somewhat unusual picture is consistent with results that were obtained using an explicit evaluation of the correlation energy in the framework of the adiabatic connection fluctuation-dissipation theorem in the random phase approximation \cite{Mittendorfer2011-PRB}. These reference calculations do not depend on a choice of exchange-correlation functional, and the agreement of our results with the reference data therefore serves as a validation of our DFT approach. 

The energy 'landscapes' associated with the two types of interface states are very different: In the physisorbed state, the energy differences between different conformations of the graphene on top of the Ni slab are minor and hence the energy 'landscape' is rather flat. In case of the chemisorbed interface, on the other hand, very significant energy changes are found when one moves the graphene between different conformations: Bridge-fcc and bridge-hcp conformations of reduced symmetry represent non-binding states of significantly increased energy, and the chemisorption minimum is also absent in the HCP-FCC configuration. This implies that it may be very difficult to 'slide' the chemisorbed graphene sheet over the Ni slab. 

\section{Molecular dynamics simulations}

We use the LAMMPS simulation package \cite{plimpton1995-J.Comput.Phys.} to perform MD simulations. We first consider planar Ni-graphene interfaces similar to our reference DFT simulations and then analyze systems consisting of a Ni coated or decorated CNT embedded into a aluminum matrix. For these simulations we need to define appropriate metal-metal, metal-carbon and C-C interaction potentials. Virtual NanoLab software \cite{vnl} (VNL Biulder) was utilized to construct intial configurations for molecular dynamics simulations.

\subsection{Interaction potentials}

In this work we consider two different sets of interaction potentials. The first type of potential  is the reactive force field (ReaxFF) which has been developed by \cite{aktulga2012-parallel,rahaman2010-J-Phy-Chem-B,shin2012-J-Phy-Chem-A,tavazza2015-J-Phy-Chem-C}. This potential is based on a bond-order model in conjunction with a charge-equilibration scheme\cite{rappe1991-J.Phys.Chem.A}. An advantage of ReaxFF is that it provides a unified description of all interactions within the ternary Al-Ni-C system and, in particular, gives an approximate representation of the chemical binding situation at chemisorbed interfaces between Ni and C. However, the potential has the drawback that it is optimized to correctly represent reactive processes rather than mechanical properties. 

\begin{figure}
\centering
	\resizebox{0.6\textwidth}{!}{%
		\includegraphics{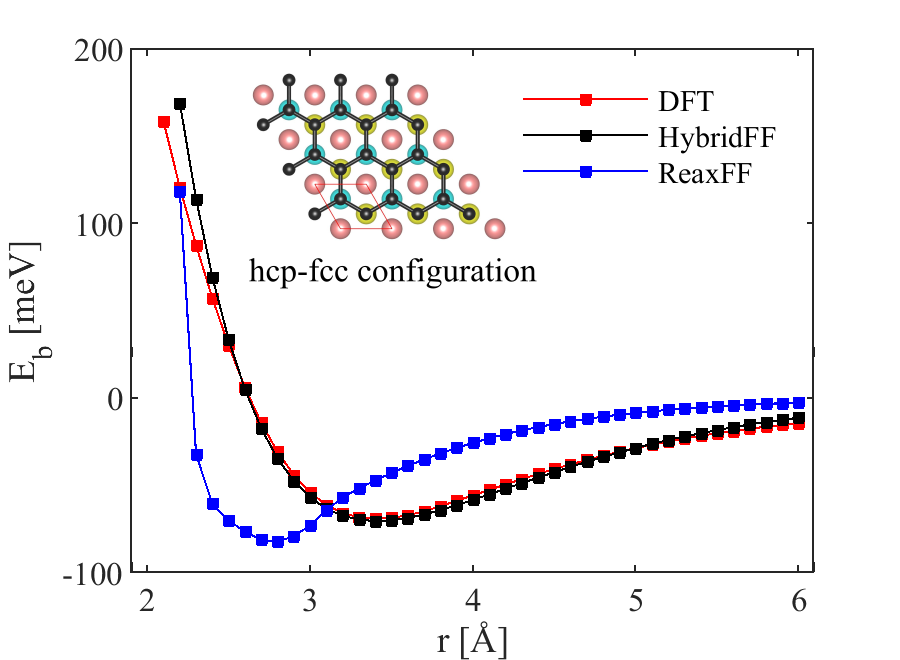} }
	\caption{Ni(111)-graphene binding energy (per carbon atom) as a function of distance; ReaxFF data, DFT data for the HCP-FCC 		configuration, and fitted Morse potential.}
	\label{fig:Morse}      
\end{figure}

A second set of potentials is based upon combining potentials of different types that are separately optimized to describe the energetics and mechanical properties of covalently bonded carbon, nickel and of aluminum. Thus we use the standard AIREBO potential \cite{brenner2002-J.Phys.Condens.Matter,stuart2000-J.Chem.Phys.} for C-C interactions and the EAM/alloy potentials of Mishin \cite{mishin2004-acta} for metal-metal interactions. The Al-C interface is described using a LJ potential. As in the approach used by \cite{Moseler2010-ACS} for simulating Ni-graphene interactions, we use a Morse potential for Ni-C which has been parametrized using our DFT calculations and fitted to represent the physisorbed interface state between graphene and Ni 
(Figure \ref{fig:Morse}). We denote this combination of potentials as HybridFF. We emphasize that such a combination of structurally dissimilar potentials can produce adequate results only as long as the interactions are additive. This is approximately true for physisorption governed by Van der Waals forces, which is not expected to significantly alter the bonding characteristics of either metals or covalently bonded carbon. On the other hand, situations where chemical bonds are established at the metal-carbon interface, or where reactive processes or mechanical mixing occur, cannot be adequately described in this manner. 

\begin{table}
	\centering
	\caption{Lattice properties of Nickel.}
	\label{tab:1}    
	\begin{tabular}{llll}
		\hline\noalign{\smallskip}
		& EamFF & ReaxFF & Experiment\cite{Lee2003-PRB} \\
		\noalign{\smallskip}\hline\noalign{\smallskip}
		\(a_{\rm 0}\) (\AA{}) &3.52&3.51&3.52\\
		\(E_{\rm 0}\) (\rm{eV})&-4.46&-4.41&-4.45\\
		$C_{11}$ (GPa) &260.7&237.7&246.5\\
		$C_{12}$ (GPa)&150.5&153.3&147.3\\
		$C_{44}$ (GPa)&131.4&153.3&124.7\\
		
	\end{tabular}
\end{table}  

\begin{table}
	\centering
	\caption{Lattice properties of Aluminum}
	\label{tab:2}    
	\begin{tabular}{llll}
		\hline\noalign{\smallskip}
		& EamFF & ReaxFF & Experiment\cite{Lee2003-PRB} \\
		\noalign{\smallskip}\hline\noalign{\smallskip}
		\(a_{\rm 0}\) (\AA{}) &4.05&4.04&4.05\\
		\(E_{\rm 0}\) (\rm{eV})&-3.36&-3.4&-3.36\\
		$C_{11}$ (GPa) &113.8&87.5&116.8\\
		$C_{12}$ (GPa)&61.5&52.3&60.1\\
		$C_{44}$ (GPa)&31.6&52.3&31.7\\
		
	\end{tabular}
\end{table}

\begin{table}
	\centering
	\caption{Lattice properties of graphene.}
	\label{tab:3}    
	\begin{tabular}{lllll}
		\hline\noalign{\smallskip}
		& ReaxFF & AIREBO & DFT\cite{spanu2009-PRL} \\
		\noalign{\smallskip}\hline\noalign{\smallskip}
		\(a_{\rm 0}\) (\AA{}) &2.46&2.46&2.45\\
		\(E_{\rm 0}\) (\rm{eV})&-7.46&-7.39&-7.46\\
		\(d_{\rm 0}\) (\AA{}) &3.29&3.41&3.35\\
		\(E_{d}\) (meV/atom C)&-68.15&-46.4&-41.5 - -58.81\\
	\end{tabular}
\end{table} 

To parametrize the Morse potential for physisorbed Ni-graphene interfaces, we use our DFT results for the interaction between Ni(111) and graphene as a reference. We use the same configuration as in the DFT simulations, i.e., we separate the graphene sheet rigidly from the Ni(111) surface, which allows us to directly compare the MD and DFT energies (Figure \ref{fig:Morse}).  The resulting parameters for the Ni-C Morse potential are \(D\) = 0.0048 eV, \(r_{\rm 0}\) = 4.18 \AA{} and $\alpha$ = 1.15 $\mathrm{\AA}^{\rm -1}$. The LJ potential used to represent the interaction between Al and C atoms is parameterized using the same configuration (hcp-fcc) as in DFT calculations by Christian et al. \cite{christian2017-carbon}. Our parameters for the LJ potential are $\epsilon$ = 0.004 eV and $\sigma$ = 3.75 \AA{}.

The performance of ReaxFF and HybridFF potentials in reproducing properties of the system components is illustrated in Table \ref{tab:1} and Table \ref{tab:2} which compare the calculated lattice constant \(a_{\rm 0}\), cohesive energy \(E_{\rm 0}\) and elastic constants of Al and Ni using ReaxFF and EAM with experimental values. Both potentials correctly represent the lattice constant and cohesive energy of Ni and Al, however, the ReaxFF cannot account for the appreciable cubic anisotropy of both materials. Lattice properties of monolayer graphene as well as the interlayer distance \(d_{\rm 0}\) and interlayer energy  \(E_{d}\) of bilayer graphene are compared in Table \ref{tab:3}. Here, both used potentials yield acceptable results. 

\subsection{Molecular simulation: Ni-Graphene interface energy surfaces}

\begin{figure}
	\centering
\resizebox{0.49\textwidth}{!}{
		\includegraphics{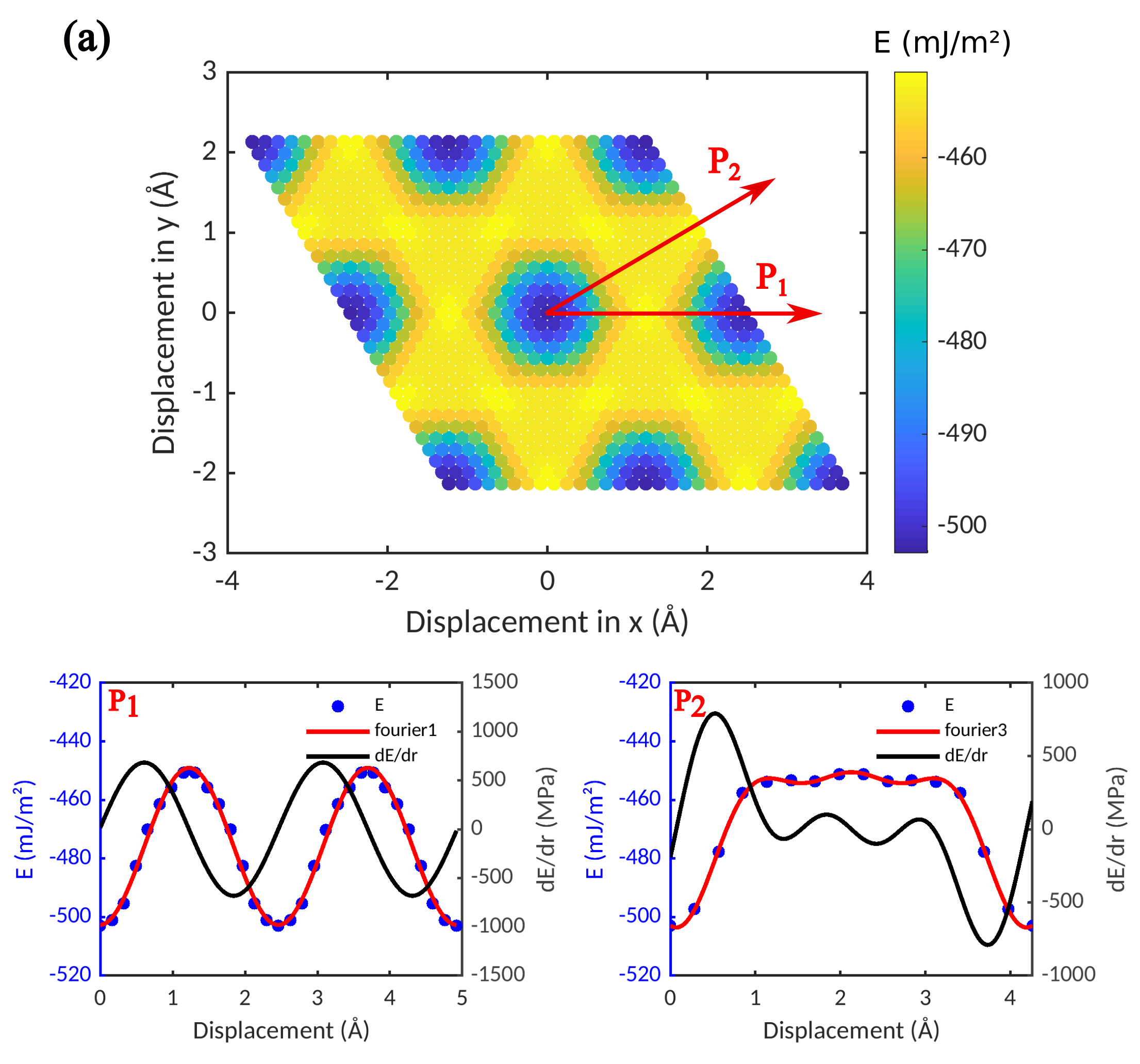}}
\resizebox{0.49\textwidth}{!}{		
		\includegraphics{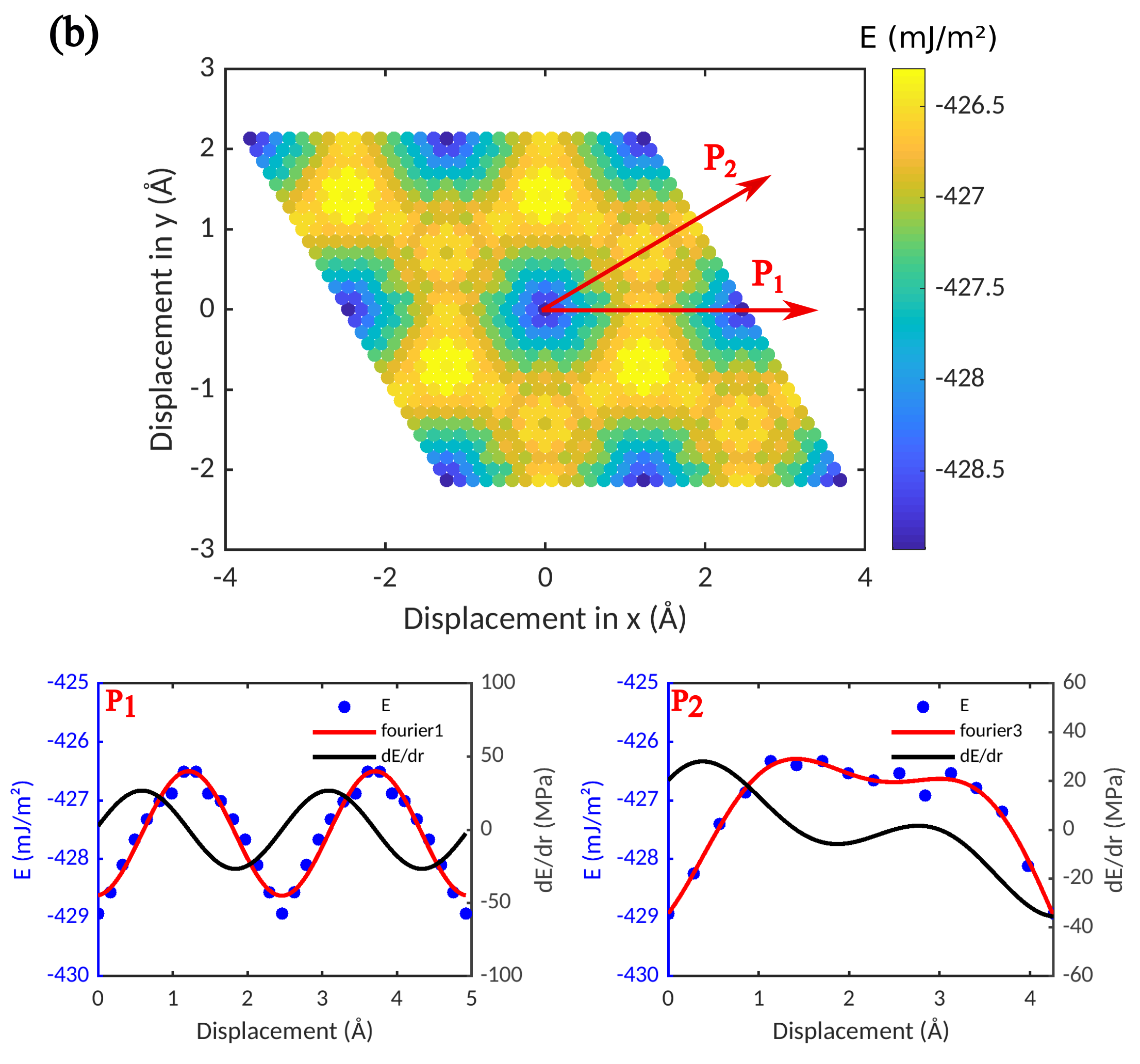}} 
	\caption{Interface energy surfaces for lateral displacement of a graphene sheet on a Ni slab, with energy and shear stress   profiles for displacement along the \(\rm P_{\rm 1}\) and  \(\rm P_{\rm 2}\) directions; left (a): ReaxFF potential, right (b): HybridFF potential; the distance between Ni and graphene is kept fixed at the distance for the respective energy minimum.}
	\label{fig:IES}       % Give a unique label
\end{figure}

We first investigate the predictions of the two potentials regarding the energy changes that occur when an adsorbed graphene sheet is rigidly displaced on top of a Ni slab. To this end, we keep the distance between Ni and graphene fixed at the value for the absolute energy minimum, and move the graphene laterally. The resulting 'interface energy surfaces' are shown in Figure \ref{fig:IES} for the HybridFF and ReaxFF potentials. 

It can be seen that both the average values of the interface energies and the overall morphology of the interface energy surfaces are comparable for the ReaxFF and HybridFF potentials. However, the interface energy variations between different configurations obtained with the ReaxFF potential exceed those obtained with the HybridFF potential by a factor of more than 10. This has important consequences for the mechanical properties of the interface. The maximum slopes of the energy vs. displacement curves along the \(\rm P_{\rm 1}\) and  \(\rm P_{\rm 2}\) directions, which define the critical shear stresses required for moving the graphene sheet along these directions, are for the ReaxFF potential about a factor of 15 higher than for the HybridFF potential. Thus, the ReaxFF potential predicts a much higher interface shear strength. Comparison of the MD results with our DFT results indicates that both HybridFF and ReaxFF provide reasonable values for the characteristic interface energy. By construction, the interface width obtained with HybridFF matches the DFT data for the physisorbed graphene sheet, whereas the Ni-graphene spacing obtained from ReaxFF lies closer to the DFT results for the chemisorbed state. Finally, the interface energy landscape obtained with ReaxFF shows much more pronounced variations than that obtained with HybridFF. This is again in qualitative agreement with DFT data, which indicate a comparatively flat interface energy landscape for the physisorbed state and much larger interface energy variations for the chemisorbed state. 

\subsection{Molecular simulation: CNT/Ni-Al systems, specimen preparation}

In the main part of this paper, we focus on the pull-out behavior of single- and multi-walled carbon nanotubes (SWCNTs and MWCNTs) embedded into an Al matrix. Our simulated samples consist of a single (pristine, Ni decorated or Ni coated) CNT embedded into a block of Al atoms. We refrain from using highly idealized molecular models, such as a complete or incomplete monolayers of Ni deposited coherently onto the CNT surface as considered by \cite{Song2010_CMS,Duan2017_PhysicaE}. Instead, we try to represent in a more realistic manner the structures that result from electroless plating of Ni onto CNT \cite{Li1997_JJAP,Kong2002_SCT} followed by melt processing to embed the coated CNT into an Al matrix. Electroless plating initially deposits discontinuous Ni clusters onto the CNT surface \cite{Li1997_JJAP,Kong2002_SCT}. This may lead to a discontinuous decoration of the CNT with various degrees of surface coverage, or the clusters may merge and reconstruct to form a continuous coating. To mimic this situation, different numbers of icosahedral Ni nanoparticles were initially placed at random locations on the CNT surface. To equilibriate the particles on the CNT surface and allow for the possible formation of a continuous coating, the Ni decorated CNT was annealed at 2300K (i.e., above the melting temperature of Ni) for 20 ps in the $NpT$ ensemble at zero pressure, and then quenched to 0.1 K at a rate of 10 K/ps. During this anneal-quench cycle the CNT atoms were kept fixed. The Ni decorated CNT were then embedded into an Al matrix. The embedding k has equal length to the embedded nanotube, which is aligned with a [110] crystal lattice direction. The lateral extension of the Al block was close to 10 nanotube diameters, and periodic boundary conditions were imposed in lateral directions. To create the requisite space, all Al atoms were removed that were located inside the CNT or within a distance of less than 3.0  $\mathrm{\AA}$ from C or Ni atoms. To allow for structural relaxation of the Al matrix around the embedded CNT, the system was then annealed at a temperature of 1500K  for 50 ps in the $NpT$ ensemble and finally cooled to 0.1 K at a rate of 10 K/ps. The temperature in this second annealing step was chosen in between the melting temperatures of the Ni coating and the Al matrix, such that the Al matrix but not the Ni coating underwent a melt-quench cycle. Within the limitations of MD, which require an unrealistically high cooling rate, this method of relaxation mimics the processing of composites fabricated via a melt process. In a last step, the energy of the system was minimized in a molecular mechanics framework through structural relaxation using a conjugate gradient algorithm. 

\begin{figure}
	\centering
	\resizebox{0.48\textwidth}{!}{%
		\includegraphics{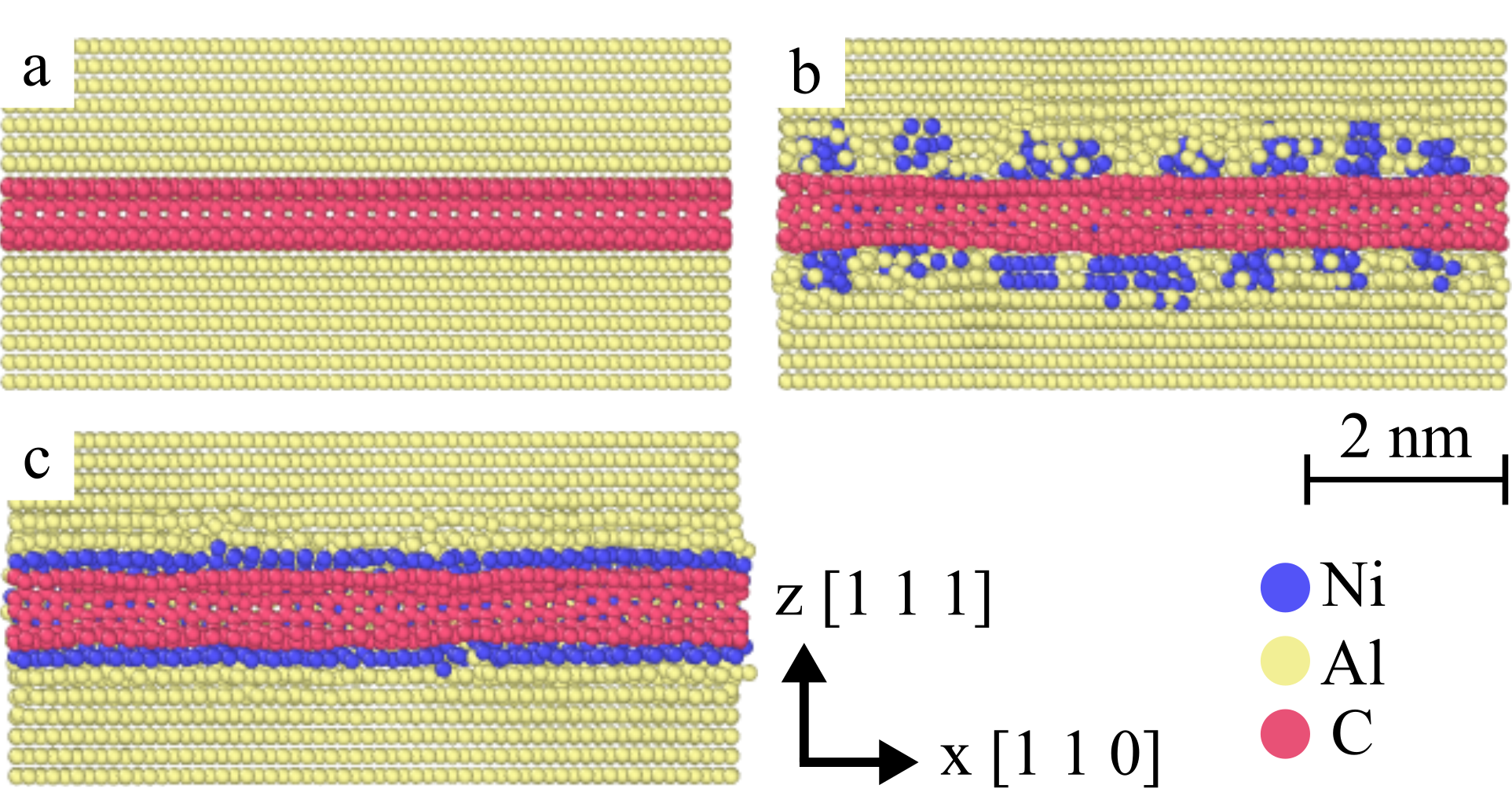} }
	\caption{CNT(5,5) surface decorated with Ni nanoclusters, (a-c) 0\%, 80\% and 100\% nickel coverage respectively.}
	\label{fig:CNTsingle}     
\end{figure}

\begin{figure}
	\centering
	\resizebox{0.48\textwidth}{!}{%
		\includegraphics{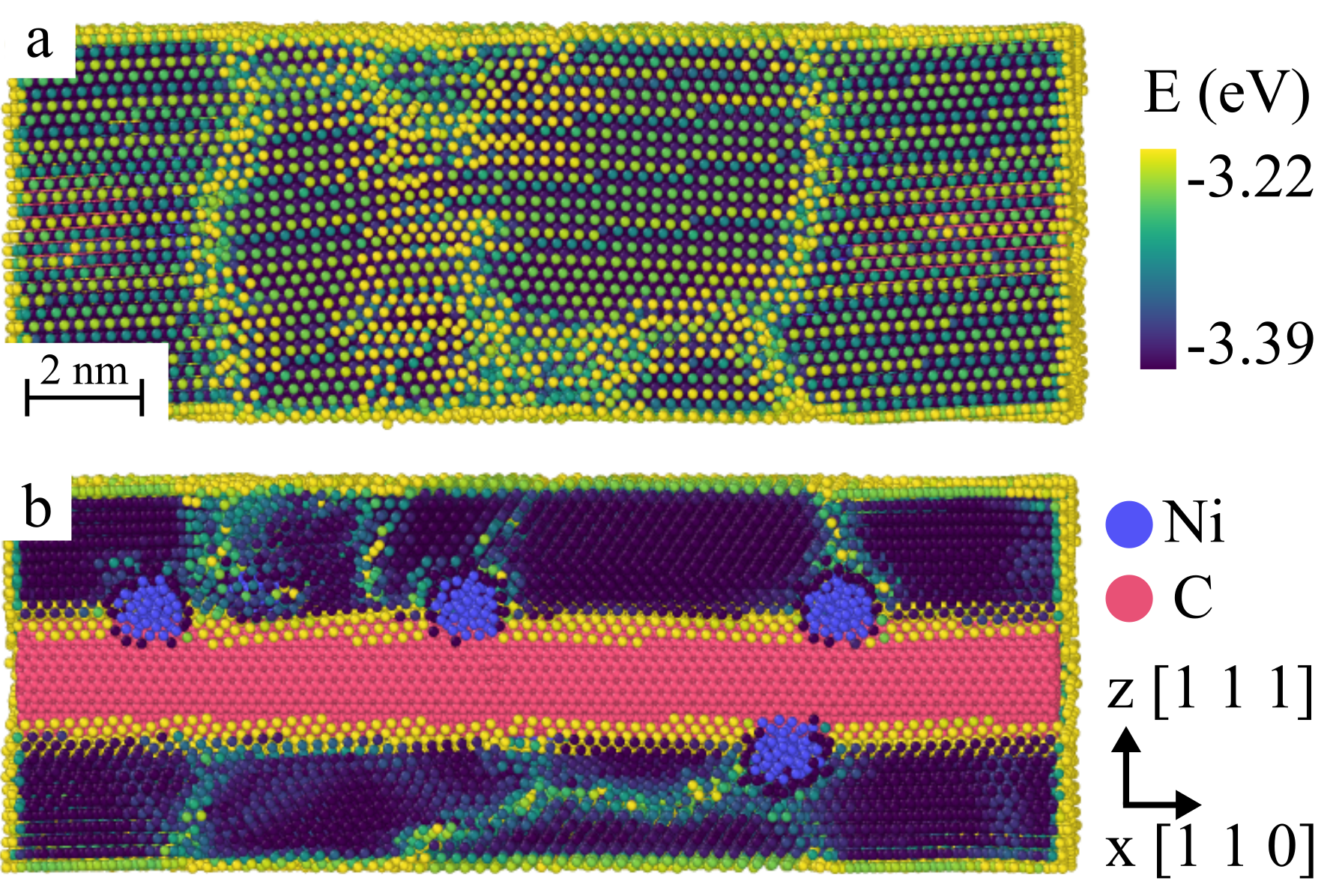} }
	\caption{Nano-crystallization of Al around a Ni decorated MWCNT; Al atoms are colored according to their potential energy in order to visualize the grain boundaries; (a): side view of simulation cell, (b): central section of simulation cell.}
	\label{fig:CNTpoly}       % Give a unique label
\end{figure}

During the melt-quench cycle at 1500 K, which relaxes the Al matrix around the embedded CNT, two different types of constraints were applied at the periodic boundaries: (i) in our standard simulation format, we kept the boundary atoms fixed on the sites of the original crystal lattice. In this case, crystallization starts from the boundary and leads to a monocrystalline Al block surrounding the CNT, which remains aligned with a [110] lattice direction as shown in Figure \ref{fig:CNTsingle}. (ii) in a number of simulations of Ni decorated MWCNTs, the boundary atoms were left free to move. In this case, crystallization starts from the Ni nanoclusters decorating the MWCNT and leads to a nanocrystalline structure. This is illuatrated in Figure \ref{fig:CNTpoly} showing the microstructure of the Al matrix surrounding a (15,15)(10,10)(5,5) MWCNT with length of 200 $\mathrm{\AA}$ and decorated with 10 Ni nano-clusters of about 10  $\mathrm{\AA}$  diameter.

We see here an interesting potential benefit of using CNT decorated by Ni nano-particles as fillers in melt processed Al matrix composites. The Ni nano-particles have a high surface energy and are therefore efficient nucleation sites for Al solidification. Thus, in melt processed Al-CNT composites, well dispersed Ni decorated CNT might not only act as fillers but also as nanocrystallization agents. The resulting nanocrystalline structure is expected to significantly modify the plastic deformation properties of the Al matrix and to enhance its resistance against plastic deformation. 

\subsection{Molecular simulation: CNT/Ni-Al systems, CNT pull-out tests}

In the remainder of this investigation, we focus on simulated pull-out testing of the embedded CNTs. During a simulated pull-out test, the velocities of Al atoms at the boundary of the simulation box were fixed to zero, and a constant velocity of 0.1 \AA/ps was imposed for 1000 ps on the CNT front layer (30 Carbon atoms). The remaining atoms were thermostatted to a temperature of 1K using a Nose-Hoover thermostat with a characteristic relaxation time of 0.1 ps. In order to have sufficient empty space in pull-out direction, the simulation box was in this direction extended to more than twice the CNT length. During pull-out the potential energy as well as the total reaction force acting on the CNT front layer were recorded. 

\begin{figure*}
\centering
	\resizebox{0.7\textwidth}{!}{%
		\includegraphics{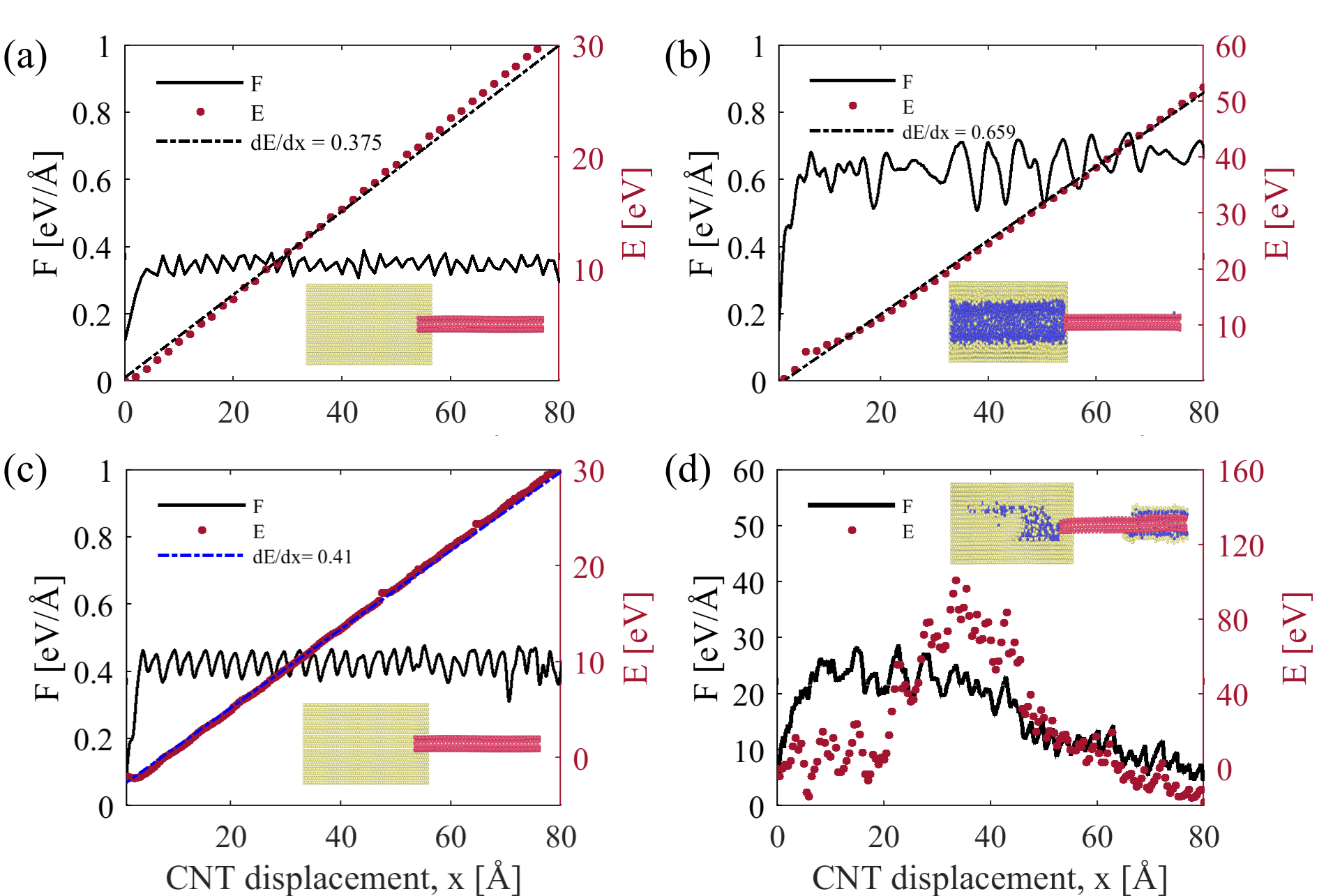} }
	\caption{Simulation of CNT pullout using HybridFF(a,b) and ReaxFF (c,d)); potential energy and pullout force as a function of CNT displacement; right: fully Ni coated CNT in Al matrix, left: pristine CNT in Al matrix.}
	\label{fig:CNTpull}       % Give a unique label
\end{figure*}

\begin{figure}
	\centering
	\resizebox{0.5\textwidth}{!}{%
	\includegraphics{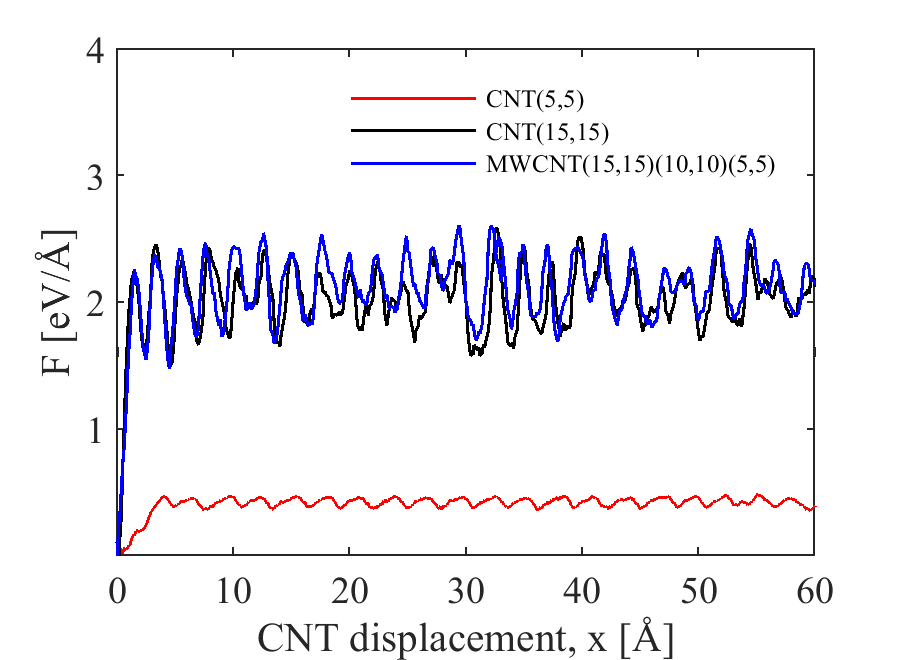}}
	\caption{Simulation of (a) pristine and (b) decorated CNT pullout using ReaxFF; (a): effect of CNT diameter and number of layers on pullout force and (b): comparison of pullout force of Ni decorated MWCNT(15,15)(10,10)(5,5) (10\% Ni coverage) from monocrystalline and nanocrystalline Al matrices. }
	\label{fig:CNTdiam}       % Give a unique label
\end{figure}

Figure \ref{fig:CNTpull} depicts the differences in pullout force and energy between pristine and fully Ni coated (5,5) SWCNT. We first consider simulations performed using the HybridFF potential as shown in Figure \ref{fig:CNTpull}(a,b). In the reference simulation which considers a pristine CNT, the pullout force oscillates around a value of 0.375 eV/$\mathrm{\AA}$ (0.64 nN), which matches the average slope of the potential energy increase and corresponds to the product of CNT circumference and interface energy. The corresponding interface shear stress between the pristine CNT and Al matrix (surface area 1858.5 \AA$^2$) is 34.5 MPa, which is in a good agreement with experimental data \cite{zhou2016-carbon}. If we now move to a fully Ni coated CNT and assume a physisorbed Ni-CNT interface as implicit in the HybridFF potential, the pullout force in the simulations fluctuates around 0.66 eV/$\mathrm{\AA}$ which is, again, equal to the derivative of potential energy with respect to CNT displacement. The modest strengthening effect in these simulations thus exclusively derives from the fact that a Ni-CNT interface has a slightly lower energy than an Al-CNT interface. At the end of the simulation, the potential energy increase of the system quite exactly matches the mechanical work expended during CNT pull-out, indicating that there is little or no energy dissipation. The conversion of mechanical work into interface energy has the following consequences: (i) the pull-out force is approximately proportional to the circumference of the CNT, with a small correction due to the curvature dependence of interface energy, (ii) the pull-out force does not depend on the length of CNT embedded into the Al matrix. In conjunction with the low binding energy these findings, if correct, would make the embedded CNT a very poor reinforcement. 

\begin{figure*}
	\centering
	\resizebox{0.99\textwidth}{!}{%
	\includegraphics{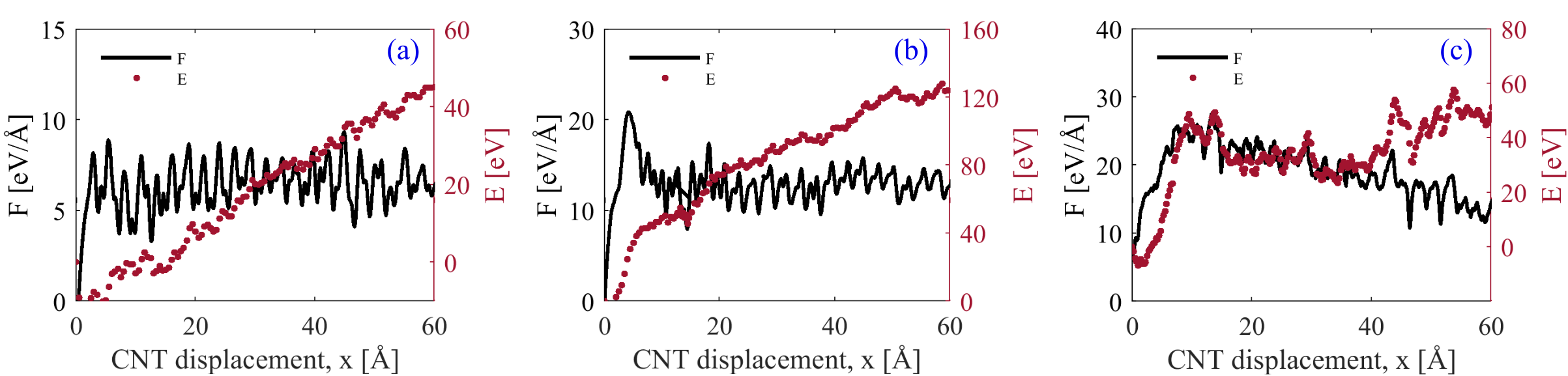} }
	\caption{Simulation of CNT(5,5) pullout using ReaxFF: effect of Ni coverage on pullout force and potential energy vs. distance curve; 
	(a): 10 \% Ni coverage, (b): 20 \% Ni coverage, (c): 40 \% Ni coverage.}
	\label{fig:CNTcover}      
\end{figure*}

Using ReaxFF to describe Ni-C interactions produces a quite different picture (Figure \ref{fig:CNTpull} (c,d)). In the reference simulations which consider a pristine CNT, the behavior is similar to the HybridFF simulations, with the only difference being a slightly higher pull-out force in line with a slightly lower energy of the  Al-CNT interface with ReaxFF interaction potential. Again, the pull-out force is approximately proportional to the CNT diameter and independent of embedded CNT length. Simulations of MWCNT indicate that inner layers have no influence on the pull-out force (Figure \ref{fig:CNTdiam} (a)). All these observations are consistent with the idea that pull-out of pristine CNT is governed exclusively by Al-CNT interface energy. Comparison of pull-out simulations from monocrystalline and nanocrystalline Al matrices also shows no significant differences in pull-out force. (Figure \ref{fig:CNTdiam} (b))

\begin{figure}
	\centering
	\resizebox{0.6\textwidth}{!}{%
	\includegraphics{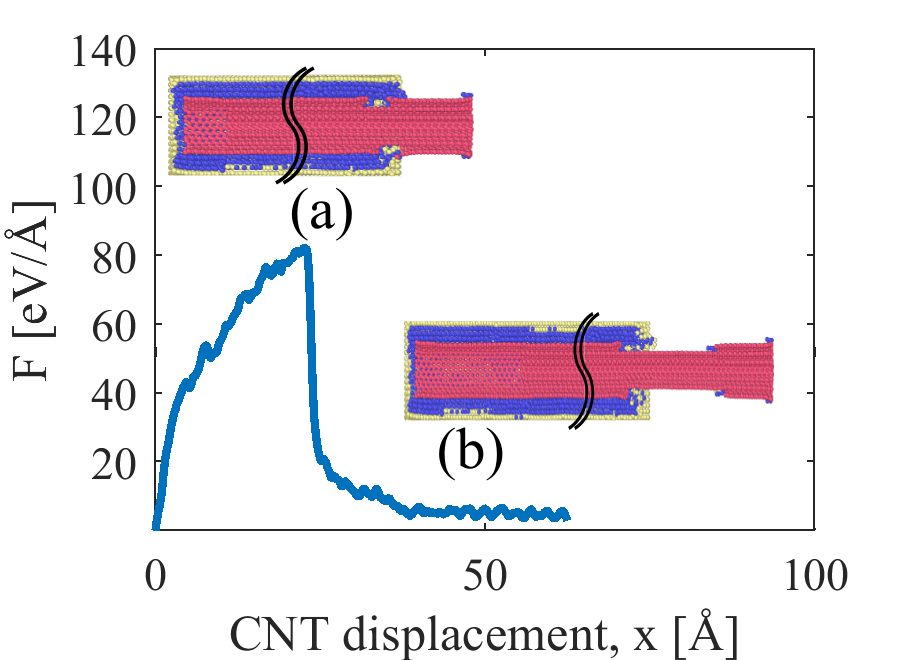} }
	\caption{Simulation of CNT pullout using ReaxFF; (a): fracture of MWCNT(15,15)(10,10)(5,5) from the outer layer, (b): sliding inner walls.}
	\label{fig:CNTbreak}       % Give a unique label
\end{figure}

However, the pull-out behavior of a Ni coated SWNT is completely different when interactions are described with ReaxFF: We find a huge pullout force up to 30 eV/\AA, which exceeds the pull-out force of the pristine CNT by almost two orders of magnitude and comes very close to the CNT fracture force. The potential energy changes during the pull-out process (it first increases and then, after rupture of the Ni coating, decreases), but these changes do not affect the pull-out force which is consistently much higher than the slope of the potential energy vs. distance curve. Globally, the pull-out process is not associated with any major increase in potential energy as the potential energies of the system before and after pull-out are almost identical. Thus, the work expended during pull-out is almost completely converted into heat and absorbed by the thermostat: the pull-out process is fully dissipative. The reasons for this behavior can be explained as follows. In the ReaxFF simulations, the large energy variations that occur when the CNT moves relative to the Ni coating make it difficult to pull the CNT out of its Ni coating. At the same time, the higher surface energy of Ni implies that it is initially energetically favorable to pull not only the Ni coating out of the Al block, but also a surrounding Al layer. This pull-out occurs by irreversible shearing of the Al matrix adjacent to the Ni coating which entails a very significant energy dissipation. Further sources of dissipation are the rupture of the Ni coating and the mechanical mixing that occurs at the Al-Ni interface during later stages of the pull-out process (see Figure \ref{fig:CNTpull} (d)). As dissipation is related to interface processes, the amount of energy dissipated per unit CNT displacement and accordingly the pull-out force decrease in approximate proportion with the length of CNT that is still contained in the Al block. 

The fact that Ni coated CNT exhibit, in our simulations, very different pull-out behavior depending on the description of the interface poses intriguing problems in view of the DFT results. We fitted the Morse potential in our HybridFF to mimic the physisorbed interface state, while the ReaxFF may be thought of reproducing the chemisorbed state. Unlike the standard picture of physisorption vs. chemisorption, both states are in case of Ni-graphene interfaces of approximately equal energy, i.e., chemisorption does not produce appreciably stronger binding. However, the dynamic response during pull-out is radically different. The chemically bonded state between graphene and Ni is characterized by an energy 'landscape' that exhibits huge variations as one moves the graphene over the Ni surface. This salient feature carries over to the more complex case of a Ni coating on top of a CNT, where the energy variations arising from the formation and breaking of chemical bonds make it difficult to pull the CNT out of its Ni coating. This observation vividly illustrates that the mere value of interface energy may be quite insufficent to explain the shear strength of interfaces. 

It remains to ask which of the two ways of describing the interface (chemisorption described by ReaxFF vs. physisorption described by HybridFF) is actually more appropriate. To this end, we consider a fully coated CNT and note that the cylindrical geometry of the system implies that a tighter Ni coating may be energetically favorable even if the interaction energy between Ni surface and CNT is the same for physisorption and chemisorption, simply because the tighter chemisorbed structure has less Ni surface around the CNT. Furthermore, a coated and embedded CNT with a chemically bonded interface cannot easily switch its interface configuration to the physisorbed configuration because of the constraint imposed by the embedding Al matrix. For a pristine CNT, on the other hand, both ReaxFF and HybridFF predict similar behavior, which is in line with the fact that Al-graphene interfaces do not exhibit chemisorption but Van der Waals bonding only. In view of these considerations, we consider in the subsequent simulations only the ReaxFF potential. 

In order to see the effect of Ni coverage on pullout force we performed pullout simulations of (5,5)SWCNTs decorated by different amounts of Ni nano-particles. Figure \ref{fig:CNTcover} shows the pullout force and potential energy change of CNTs with 10\%, 20\%, and 40\% Ni surface coverage. The pull-out force increases with increasing Ni surface coverage and, in all cases, the overall work expended during CNT pull-out is much higher than the potential energy increase. The process is thus mainly dissipative even at the lowest degree of Ni coverage. At the same time, certain changes occur as surface coverage is increased.
For 10\% and 20\% of surface coverage by Ni nanoclusters, the CNT is pulled out of the Ni nanoclusters that are attached to its surface. This leads to a monotonic increase of potential energy. At the same time, at the rear end of the CNT a significant amount of mechanical mixing occurs. The energy dissipation associated with this mixing controls the overall pull-out force, which is independent of the CNT length still contained in the Al matrix. At higher degrees of Ni coverage (40\% and above) the picture changes and becomes similar to the pull-out behavior of a fully coated CNT. In this case, the coating is partially dragged through the surrounding Al matrix, the pull-out force becomes independent of surface coverage, and it decreases as the length of CNT still contained in the Al block decreases. 

The high pull-out forces of coated CNT imply that CNT rupture may occur. Figure \ref{fig:CNTbreak} shows a simulation  where, at a critical pulling force, the outer layer of an embedded Ni-coated MWCNT ruptures. Because of the weak Van der Waals bonding between adjacent walls, this leads to a failure mode where outermost-wall fracture is followed by inner wall sliding. This behavior corresponds to the sword-in-sheath fracture mechanism which has been observed during in-situ tensile tests of MWCNTs \cite{chen2015-CSTE}.

\section{Conclusions}

In this study we performed a multiscale simulation study in view of understanding the pull-out behavior of Ni decorated carbon nanotubes from an Al matrix. We find that the physical nature of the interactions between metal and CNT has a very significant influence on the pull-out force. In case of Al-CNT interfaces, Van der Waals interactions prevail which lead not only to weak bonding but also to a rather 'flat' interface energy landscape, which facilitates pull-out. For CNT coated with Ni, 
we performed DFT calculations which use a graphene layer on a Ni slab as a proxy system. In this case, the Ni-graphene interface shows two different states of approximately equal energy: a physisorbed state with Van der Waals-like bonding and a chemisorbed state which, while energetically almost equivalent, is characterized by directed bonds and a much more 'corrugated' interface energy landscape. We then prepared specimen for CNT pull-out simulations, taking care to mimic the processes that occur during preparation of real-world composites by electroless CNT plating followed by melt processing. Using a ReaxFF force field to describe the chemical interactions between Ni and C atoms, our simulations demonstrate that pull-out of Ni coated CNTs may be governed by dissipative processes, which lead to a high pull-out force that may exceed the force for CNT rupture. These findings are in qualitative agreement with previous studies \cite{Song2010_CMS,Duan2017_PhysicaE} which use different potentials to describe the C-Ni-Al system and which consider much more idealized initial configurations. They confirm that Ni coating of CNT may very substantially increase both pull-out force and energy dissipation during CNT pull-out. These factors are key prerequisites for toughening composites through CNT acting as crack bridges. Our simulations also show that CNT decorated by isolated Ni clusters may provide additional benefits in melt processed Al matrix composites. During solidification, the Ni nano-clusters might act as nucleation sites for crystallization of the Al matrix, leading to a nanocrystalline structure with enhanced resistance to plastic deformation. 

The present study has clear limitations. These result from the need to adequately consider chemical changes at the Ni-carbon interface, and at the same time to correctly capture the mechanical behavior of both CNT and metals. Combining EAM potentials for metal-metal interactions with AIREBO or Brenner-type potentials for carbon \cite{Song2010_CMS,Duan2017_PhysicaE} may provide an excellent description of the mechanical properties of the metal matrix and the CNT, and works well for composite systems as long as the interactions between CNT and metals are of Van der Waals type and can be simply added to the metallic and covalent interactions describing the constituents of the composite. If, however, chemical bonds develop at the interface between metal and CNT, as demonstrated by our DFT calculation, then it is not trivial to assume that the ensuing modifications of the electronic structure leave the interactions of carbon-bonded metal atoms with other atoms of the metal matrix unaffected. In the present study we circumvent this problem by using a ReaxFF potential specifically designed for describing changes in chemical bonding among C, Ni and Al atoms. This potential provides a unified framework that adequately describes the differences in bonding between Al and graphene, and Ni and graphene, and which captures the chemical bonding at the Ni-Carbon interface. However, this advantage comes at a price in form of a less accurate description of the mechanical properties of the Al matrix and of the Ni-Al alloy system. Further investigations are therefore needed to develop a fully satisfactory model of the complex chemo-mechanical system formed when CNT are coated by Ni and then embedded into Al. Even with its present limitations, our investigation vividly illustrates that the mechanical properties of interfaces may in many cases be inadequately captured by computational models that correctly reproduce interface energies but fail to account for specific bonding structure. In particular, we think that our findings may act as a caveat against an uncritical use of Van der Waals like interaction potentials (Morse or VdW), which produce flat interface energy landscapes and thereby may significantly under-estimate the resistance of interfaces against sliding.

\section{Author contributions}
SN performed MD simulations and analyzed MD data, KW and MJY performed DFT simulations. QL provided unpublished experimental data regarding mechanical properties, CNT coating, and solidification behavior of Al-CNT composites which guided the simulations. MZ is responsible for the simulation design.  

\section*{Acknowledgements} SN, KW, MJY and MZ acknowledge support of this research from DFG under grant no. 1 Za 171/11-1. SN and MZ also acknowledge support from DFG through the EAM Cluster of Excellence, project EXC 315. The authors gratefully acknowledge the compute resources and support provided by the Erlangen Regional Computing Center (RRZE).


\section*{References}

\begin{thebibliography}{10}

\bibitem{demczyk2002-J.Mater.Sci.Eng.A}

Demczyk, B.G., Wang, Y.M., Cumings, J., Hetman, M., Han, W., Zettl, A., Ritchie, R.O., J. Mater. Sci. Eng. A. \textbf{334}, 173 (2002)

\bibitem{yu2000-Science}

Yu, M-F., Lourie, O., Dyer, M.J., Moloni, K., Kelly, T.F., Ruoff, R.S., Science \textbf{287}, 637 (2000)

\bibitem{Nasiri2016}

Nasiri, S., Zaiser, M., AIMS Mater. Sci. \textbf{3}, 1340 (2016)

\bibitem{li2009-compos.sci.technol}

Li, Q., Viereckl, A., Rottmair, C. A., Singer, R.F., Compos. Sci. Technol. \textbf{69}, 1193 (2009)

\bibitem{george2005-Scripta.Materialia}

George, R., Kashyap, K., Rahul, R., Yamdagni, S., Scr. Mater. \textbf{53}, 1159 (2005)

\bibitem{bakshi2010-Int.mater.rev}

Bakshi, S.R., Lahiri, D. , Agarwal, A., Int. Mater. Rev. \textbf{55}, 41 (2010)


\bibitem{li2010-compos.sci.technol}

Li, Q., Rottmair, C. A., Singer, R. F., Compos. Sci. Technol. \textbf{70}, 2242 (2010)

\bibitem{kim2009-Synthetic.Metals}

Kim, C., Lim, B., Kim, B., Shim, U., Oh, S., Sung, B., Choi, J., Ki, J., Baik, S., Synth. Met. \textbf{159}, 424 (2009)

\bibitem{guo2002-J.Mater.Sci.Eng.A}

Guo, M.T., Tsao, C., J. Mater. Sci. Eng. A. \textbf{333}, 134 (2002)

\bibitem{si2008-chem.of.Mat}

Si, Y., Samulski, E.T., Chem. Mater. \textbf{20}, 6792 (2008)

\bibitem{Silvestre2014_CST}

Silvestre, N., Faria, B., Lopes, J.N.C., Compos. Sci. Technol. \textbf{90}, 16 (2014)

\bibitem{Choi2016_CB}

Choi, B.K., Yoon, G.H. , Lee, S., Compos. Part B-Eng \textbf{91}, 119 (2016)

\bibitem{Song2010_CMS}

Song, H.Y., Zha, X.W., Comput. Mater. Sci. \textbf{49} 899 (2010)

\bibitem{Duan2017_PhysicaE}

Duan, K., Li, L., Hu, Y., Wang, X., Physica E \textbf{88}, 259 (2017)

\bibitem{Shibuta2007_CMS}

Shibuta, Y. , Maruyama, S., Comput. Mater. Sci. \textbf{39}, 842 (2007)

\bibitem{giannozzi2009-J-Phy-Con-Mat}

Giannozzi, P., Baroni, S., Bonini, N., Calandra, M., Car, R., Cavazzoni, C., Ceresoli, D., Chiarotti, G.L., Cococcioni, M., Dabo, I. and others, J. Phys. Condens. Matter. \textbf{21}, 395502 (2009) 

\bibitem{blochl1994-PRB}

Blöchl, P.E., Phys. Rev. B. \textbf{50}, 17953 (1994)

\bibitem{graziano2012-J-Phy-Con-Mat}

Graziano, G., Klimeš, J., Fernandez-Alonso, F., Michaelides, A., J. Phys. Condens. Matter. \textbf{24}, 424216 (2012)

\bibitem{silvestrelli2015-PRB}

Silvestrelli, P.L., Ambrosetti, A., Phys. Rev. B. \textbf{91}, 195405 (2015)

\bibitem{munoz2016-Theory-Chem-Accounts}

Muñoz-Galán , H., Viñes, F., Gebhardt, J., Görling, A. , Illas, F., Theor. Chem. Acc. \textbf{135}, 165 (2016)

\bibitem{becke1988-PRA}

Becke, A.D., Phys. Rev. A. \textbf{38}, 3098 (1998)

\bibitem{Mittendorfer2011-PRB}

Mittendorfer, F., Garhofer, A., Redinger, J., Klimeš, J., Harlm, J., Kresse, G., Phys. Rev. B. \textbf{84}, 201401 (2011)

\bibitem{plimpton1995-J.Comput.Phys.}

Plimpton, S., J. Comput. Phys. \textbf{117}, 1 (1995)

\bibitem{vnl}

Virtual Nanolab version 2017.2 QuantumWise A/S (www.quantumwise.com)

\bibitem{aktulga2012-parallel}

Aktulga, H.M., Fogarty, J.C., Pandit, S.A., Grama, A.Y., Parallel Comput. \textbf{338} 245 (2012)

\bibitem{rahaman2010-J-Phy-Chem-B}

Rahaman, O., Van Duin, A.C.T., Goddard III, W.A., Doren, D.J., J. Phys. Chem. B \textbf{115}, 249 (2010)

\bibitem{shin2012-J-Phy-Chem-A}

Shin, Y.K., Kwak, H., Zou, C. Vasenkov, A.V., Van Duin A.C.T., J.Phys. Chem. A. \textbf{116}, 12163 (2012)

\bibitem{tavazza2015-J-Phy-Chem-C}

Tavazza, F., Senftle, T.P., Zou, C., Becker, C.A., van Duin, A.C.T., J. Phys. Chem. C. \textbf{119}, 13580 (2015)

\bibitem{rappe1991-J.Phys.Chem.A}
Rappe, A.K., Goddard III, W.A., J. Phys.Chem. \textbf{95}, 3358 (1991)

\bibitem{brenner2002-J.Phys.Condens.Matter}

Brenner, D.W., Shenderova, O.A., Harrison, J.A., Stuart, S.J., Ni, B., Sinnott, S.B., J. Phys. Condens. Matter \textbf{14}, 783 (2002)

\bibitem{stuart2000-J.Chem.Phys.}

Stuart, S.J., Tutein, A.B., Harrison, J.A., J. Chem. Phys. \textbf{112}, 6472 (2000)

\bibitem{mishin2004-acta}

Mishin, Y., Acta. Mater. \textbf{52}, 1451 (2004)

\bibitem{Moseler2010-ACS}

Moseler, M., Cervantes-Sodi, F., Hofmann, S., Csányi, G., Ferrari, A.C., ACS Nano. \textbf{4}, 7587 (2010)


\bibitem{Lee2003-PRB}

Lee, B.J., Shim, J.H. , Baskes, M.I., Phys. Rev. B. \textbf{68}, 144112 (2003)

\bibitem{spanu2009-PRL}

Spanu, L., Sorella, S. , Galli, G., Phys. Rev. Lett. \textbf{103}, 196401 (2009)

\bibitem{christian2017-carbon}

Christian, M.S., Otero-de-la-Roza, A., Johnson, E.R., Carbon \textbf{124}, 531 (2017)	
\bibitem{Li1997_JJAP}

Li, Q., Fan, S., Han, W., Sun, C., Liang, W., Jpn. J. Appl. Phys. \textbf{36}, 501 (1997)

\bibitem{Kong2002_SCT}

Kong, F.Z., Zhang, X.B., Xiong, W.Q., Liu, F., Huang, W.Z., Sun, Y.L., Tu, J.P., Chen, X.W., Surf. Coat. Technol. \textbf{155}, 33 (2002)

\bibitem{zhou2016-carbon}

Zhou, W., Yamamoto, G., Fan, Y., Kwon, H., Hashida, T., Kawasaki, A., Carbon \textbf{106}, 37 (2016)

\bibitem{chen2015-CSTE}

Chen, B., Li, S., Imai, H., Jia, L., Umeda, J., Takahashi, M., Kondoh, K., Compos. Sci .Technol. \textbf{113}, 1 (2015)
	
	
\end{thebibliography}
\end{document}